\documentstyle[epsf,12pt,aaspp4,flushrt]{article}

\begin{document}
\baselineskip=12pt

{\small Astronomy Letters 1999, 25}

\title{Long-Term Variability of the Hard X-ray Source GRS~1758--258:
GRANAT/SIGMA Observations}

\author{S.I.~Kuznetsov$^{1,2}$, M.R.~Gilfanov$^{3,1}$,
E.M.~Churazov$^{3,1}$, R.A.~Sunyaev$^{3,1}$, A.V.~Dyachkov$^{1}$,
N.G.~Khavenson$^{1}$, B.S.~Novikov$^{1}$, R.S.~Kremnev$^{1}$,
P.~Goldoni$^{4}$, A.~Goldwurm$^{4}$, P.~Laurent$^{4}$, J.~Paul$^{4}$,
J.P.~Roques$^{5}$, E.~Jourdain$^{5}$, L.~Bouchet$^{5}$, G.~Vedrenne$^{5}$}

\affil{\it ~ \nl
$^{1}$ Space Research Institute, RAS, Moscow \nl
$^{2}$ Moscow Physical-Technical Institute, Dolgoprudny, Russia \nl
$^{3}$ MPI f\"ur Astrophysik, Garching, Germany \nl
$^{4}$ Service d`Astrophysique, DAPNIA/DSM, CEA Saclay, France \nl
$^{5}$ Centre d`Etude Spatiale des Rayonnements (CNRS/UPS), Toulouse Cedex,
France}

\vspace*{1cm}

{\small ~ \hfill Received November 5, 1998; in final form, December 1, 1998.}

\section*{Abstract}

The results of GRANAT/SIGMA hard X-ray observations of GRS~1758--258 in
1990--1998 are presented. The source lies at $\sim 5\arcdeg$\ from the
Galactic Center and was within the SIGMA field of view during the GRANAT
surveys of this region.  The total exposure time of the Galactic Center was
$11 \times 10^{6}$\ s.  The regular SIGMA observations revealed strong
variability of the source: the 40--150 keV flux varied at least by a factor
of 8 on a time scale of a year, between less than 13 mCrab and $\approx$\
100--110 mCrab. The average flux was $\approx$\ 60 mCrab in 1990--1998. The
source's spectrum is well fitted by a power law with a photon index $\alpha
\approx$\ 1.86 in the energy range 40 to 150 keV and becomes steeper at
energies above $\sim$\ 100 keV.  The radio and hard X-ray properties of
GRS~1758--258 are similar to those of another Galactic Center source,
1E1740.7--2942. GRS~1758--258 and 1E1740.7--2942 are the two brightest hard
X-ray sources in the Galactic Center region. Both sources have radio jets,
similar X-ray luminosities ($\sim~10^{37}$\ erg/s), and spectra, and exhibit
variations in the hard X-ray flux on long times scales by a factor of
$\sim~10$\ or more .  In contrast to most of the known black hole
candidates, which are X-ray transients, GRS~1758--258 and 1E1740.7--2942
were detected by SIGMA during most of the observations in 1990--1998.
Assuming that this behavior of the sources implies the suppression of
accretion-disk instability in the region of partial hydrogen ionization
through X-ray heating, we impose constraints on the mass of the optical
companion and on the orbital period of the binary system.

\section{Introduction}
GRS~1758--258 was discovered by the ART-P and SIGMA telescopes onboard the
GRANAT Observatory during the first series of observations of the Galactic
Center region in March--April 1990 (Sunyaev et al. 1991; Gilfanov et al.
1993). Together with 1E1740.7--2942, this source dominated in the Galactic
Center region in the 40--300 keV energy band.  The hard X-ray spectra of
GRS~1758--258 and 1E1740.7--2942, which extend up to $\sim$\ 100--300 keV,
are similar to the low-state spectra of Cygnus X--1, a well-known black hole
candidate. Their spectral properties suggest that the two sources
(GRS~1758--258 and 1E1740.7--2942) are black hole candidates (Sunyaev et al.
1991).

An analysis of the ROSAT soft X-ray observations in the Spring of 1993
led Mereghetti et al. (1994) to conclude that there is a soft X-ray
component in the spectrum of GRS~1758--258. This conclusion was called
into question by Grebenev et al. (1996), who performed an independent
analysis of these and earlier (the Spring of 1992) ROSAT observations.
Mereghetti et al. (1992) refined the source position to within 10\arcsec
(90\% error circle radius): $\alpha = 17^{h}58^{m}7.2^{s}$; $\delta =
-25\arcdeg\ 44\arcmin\ 28\arcsec$\ (epoch 1950). A comparison of the ROSAT
and GRANAT/SIGMA observations revealed a possible anticorrelation between
the soft and hard spectral components (Mereghetti et al. 1994).

The XTE observations in the standard X-ray band show that the
Power Density Spectra (PDS) of GRS~1758--258 and 1E1740.7--2942 are
similar (Smith et al. 1997) and close to that of Cygnus X--1 (Smith et
al. 1997; Miyamoto et al. 1992). X-ray flux QPOs were also found in
the PDS of the two sources (Smith et al. 1997).

VLA radio observations of GRS~1758--258 at $\lambda=6$\ cm were
carried out in early and mid--1992, when the hard X-ray flux (40--150 keV)
from the source was below the SIGMA detection threshold ($\sim$\ 13 mCrab --
$3\sigma$\ upper limit; Gilfanov et al. 1993). A triple radio source
was identified with GRS~1758--258 (Rodriguez et al. 1992). The position
of the central core coincided with the ROSAT position of the X-ray
source (Mereghetti et al.  1992), while the southern and northern
lobes were, respectively, at $\approx$\ 1\farcm 3 and $\approx$\
2\farcm 1 on either side of the core. A similar radio pattern was previously
observed in 1E1740.7--2942 (Mirabel et al. 1992; Mirabel et al. 1993).

\section{Instrument and observations}
The SIGMA telescope is one of the principal instruments onboard the
GRANAT Astrophysical Observatory. It is designed to obtain
hard X-rays and soft $\gamma$-ray (35--1300 keV) images using the 
principle of a coded mask. The angular size of the full coding region is 
4\fdg 7 $\times$\ 4\fdg 3, and the field of view at half sensitivity is
10\fdg 5 $\times$\ 10\fdg 6.
The nominal angular resolution of the telescope is 13\arcmin\' (the pixel size 
of the coded mask). Bright point sources can be localized to within 
several tens of arcseconds, depending on the source brightness and the 
number of observations. A more detailed description of the telescope was 
given by Paul et al. (1991).

Inflight energy calibrations of the telescope were performed using
observations of Crab (the last observation was in October
1997). The angular resolution was calibrated using observations of
bright compact sources (Cyg X--1, Crab , bright X-ray
transients). Correction for the background illumination of the detector
was made using observations of ``empty fields'' (regions of the sky
without bright sources).

Observations\footnote{For the light curve all 1990-1998 data were used. The
latest data (since the Fall of 1997) may be subject to slightly larger
systematic errors that those for the previous years due to the slow
evolution of the detector background illumination, whose last calibrations
were performed in September 1997. A subsequent analysis of these data can
result in small changes in the flux for the last three points in Fig.1. For
the same reason, we do not provide the best-fit parameters for these
observations in Tables 1--4.} of the Galactic Center region were carried out
yearly in the Spring and Fall of 1990--1994 and 1997--1998, in the Fall of
1995, and in the Spring of 1996, with a total duration of each individual
survey between one week and two months. The mean duration of a single
observing session was $\sim$ 20 h. The SIGMA total exposure time for 171
observations during 16 surveys of the Galactic Center region is $\sim$\ 3100
h.

\section{Light Curve}

The light curve of the source averaged over the observational series is
shown in Fig. 1. A maximum (40--150 keV) flux of $\approx$\ 90--110 mCrab was
recorded in 1990 and 1997. The observations in the Spring of 1991
revealed a decline in the flux to $\approx$\ 70 mCrab.

During the observations in August--September 1991, the intensity of the
hard X-ray (40--150 keV) emission from GRS~1758--258 was considerably
lower and did not exceed the $2\sigma$\ detection threshold in each
individual session and for the sum of all observations of this series.
The upper limit on the source flux was 14 mCrab (3 $\sigma$; Cordier et
al. 1991).

GRS~1758--258 was also below the detection limit during the subsequent
series of observations in the Spring of 1992. A statistically
significant flux from the source with a 3 $\sigma$\ upper limit of 9 mCrab
cannot be detected even by combining the data for the Fall of 1991 and
the Spring of 1992 (Gilfanov et al. 1993).

GRS~1758--258 was again detected during the following series of Galactic
Center observation in the Fall of 1992 (see Fig.1) with a flux of $\approx$\
40 mCrab. Subsequent observations showed that the source intensity continued
to vary several-fold on long time scales. The flux gradually increased from
the Spring of 1992 through the Spring of 1994, when GRS 1758--258 was again
at its maximum with a flux of $\approx$\ 80 mCrab.  During the next surveys
in the Fall of 1994 and 1995 and in the Spring of 1996 and 1997, the flux
was comparable to its mean value of $\approx$\ 50 mCrab. The source
brightness was again at a maximum in the Fall of 1997 -- the flux reached
$\approx$\ 110 mCrab and was greater than its value in the previous series
of observations in the Spring of 1997 by a factor of $\sim$\ 3. The source
also remained bright in the next two series ($\approx$\ 90 mCrab).

Note that smooth variations in the hard X-ray flux from GRS~1758--258 can
be clearly traced on long time scales, while flux variations on time
scales of several days are not always traceable. Variations in the
source (40--150 keV) flux by several times on such a time scale (a few
days) were reliably detected for the series of observations in
February--April 1991, February--April 1993 and March 1996. It should be
noted that the measurement error in the flux from GRS~1758--258 per
observing session (a mean duration of $\approx 20$\ h) is typically $\sim$\
25 mCrab at a typical flux from the source of $\sim$\ 60 mCrab.

\section{Spectral Variability}
To search for spectral variability, we grouped all SIGMA observations
of GRS~1758--258 in two ways:

\begin{enumerate}
\item Grouping of the data in observational series to study possible
spectral changes on long time scales.
\item Grouping of the data in flux to search for a possible correlation
between the spectral shape and the source flux.
\end{enumerate}

\subsection{Averaging over the Observational Series}

To study the spectral properties of the source on long time scales, we
grouped the data in observational series. The derived spectra were fitted by
the following models: a power law (Table 1), bremsstrahlung of an optically
thin plasma (Kellogg et al. 1975; Table 2), and a Comptonized disk (Sunyaev
and Titarchuk 1980; Table 3). As can be seen from the $\chi^{2}$\ values in
Tables 1--3, the last two models fit the source spectrum considerably better
than does the power law. This is undoubtedly attributable to a gradual
steepening of the spectrum at energies above $\sim$\ 100 keV. Both the
bremsstrahlung model and the Comptonized disk model provide equally
acceptable (in terms of the $\chi^{2}$\ test) fits to the observed spectra.
Note that for the assumed (in the standard theory of disk accretion)
parameters of the region of X-ray generation (Shakura and Sunyaev 1973;
Shapiro et al. 1976), the bremsstrahlung model may not be relevant.
Nevertheless, we use this model as a simple analytic fit to the spectrum.
The temperature in this model provides a convenient means of characterizing
the spectral hardness by a single parameter.

Figure 2 shows a plot of spectral hardness versus time which we obtained
by using the bremsstrahlung model. There are no hardness data for the
observations in the Fall of 1991 and the Spring of 1992, because the
flux from the source was below the SIGMA detection threshold during
these periods (see above). Note that after the source was again detected
in the Fall of 1992, no significant changes occurred in the source
spectrum -- the characteristic temperature of the bremsstrahlung
spectrum was $\sim$\ 130 keV.

We attempted to trace the possible change in spectral hardness as a
function of hard X-ray luminosity in the band 40--200 keV by using the
average spectra (Fig. 3). Figure 3 shows that as the luminosity changes
from $\sim 5 \times 10^{36}$\ erg/s to $\sim 12 \times 10^{36}$ erg/s (for
an assumed distance to the source of 8.5 kpc), the hardness remains constant
(within the error limits).

Thus, we found no reliable evidence for variations in the source
spectral hardness from one observational series to another.

\subsection{Grouping in Flux}

To search for spectral variability which correlates with variations in
the source flux on time scales of days, we grouped the data in flux. A
procedure similar to that described below was used to analyze the
observations of Cygnus X-1 and 1E1740.7--2942 (Kuznetsov et al. 1997)
and revealed a clear correlation between the flux and the spectral
hardness. Since the statistical significance of the observations of GRS
1758--258 is lower, we cannot perform a detailed analysis of the
flux-hardness correlation. For completeness, however, we carried out
such an analysis.

The data were averaged as follows. The entire range of variations in the
40--150 keV flux observed in 1990--1997 was divided into several bins of equal
width. Accordingly, all individual observing sessions (in which the source
flux exceeded 1.4 $\times 10^{-3}$\ counts sec$^{-1}$ cm$^{-2}$) were
divided into groups; sessions in which the flux was in a given bin fell in
each group. We then averaged the source spectra in each group of sessions.

As a result, we obtained average spectra of the source for various flux
ranges. The average spectra were fitted by the model of bremsstrahlung of an
optically thin plasma in the band 40--300 keV. For each spectrum, we
determined the best-fit luminosity and hardness ($kT$). The data covered the
luminosity range from $\sim 5 \times 10^{36}$\ erg/s to $\sim 16 \times
10^{36}$\ erg/s (for an assuming distance to the source of 8.5 kpc). The
results are shown in Fig. 4. We see that the spectral hardness ($kT$) is
constant (within the error limits): the mean temperature of the
bremsstrahlung spectrum is $kT \approx$\ 130 keV. The previously found
correlation ``spectral hardness -- hard X-ray luminosity'' for Cygnus X--1
and 1E1740.7--2942 (Kuznetsov et al. 1997), which is indicated by the dashed
line in Fig. 4, can be neither confirmed nor rejected with a sufficient
statistical significance for GRS~1758--258.

\section{Average Spectrum of the Source}

The total exposure time for all SIGMA observations of the source in
1990--1997 was $9.3 \times 10^{6}$ s (except for the observations since the
Fall of 1997; see the footnote to Sec. 2). Assuming that the spectral shape
did not change during this period, we averaged all the source spectrum that
were obtained during the observations in 1990--1997. The average energy
spectrum (40--300 keV) was fitted by the following models: a power law, a
power law with an exponential cutoff at high energies, optically--thin
thermal bremsstrahlung (Kellogg et al. 1975), and a Comptonized disk
(Sunyaev \& Titarchuk 1980). The results are given in Table 4 and shown in
Fig. 5.

The average spectrum is well described by the models with a cutoff at
high energies -- the Comptonized disk model, a power law with an
exponential cutoff, and the bremsstrahlung spectrum: $\chi^2 =
44.9/45.1/46.0$ d.o.f.= 43/43/44, respectively (see Table 4). As we 
see from Fig. 5, the spectrum can be accurately described by a power law
only up to energies $\sim$\ 100--200 keV. At higher energies, the spectrum
steepens appreciably. Fitting of the spectrum by a power law in the band
40--300 keV gives a high $\chi^2$\ value of 59.9 (per 44 d.o.f.).

Best-fit parameters for the spectrum of 1E1740.7--2942 which was
averaged over the same set of observations are also given in Table 4. As
can be seen from this table, the best-fit parameters for the spectra of
1E1740.7--2942 and GRS~1758--258 are in good agreement. The difference
between the mean 40--200 keV luminosities is $\approx$\ 20--25 \% assuming
the same distances to the sources).

\section{Discussion}

Thus, the observational data suggest that 1E1740.7--2942 and GRS
1758--258, the two brightest (at energies above 35 keV) sources within a
few degrees of the Galactic Center, are similar. A total of six bright
sources (1E1740.7--2942, GRS~1758--258, GX~1+4, GX~354--0, 4U~1722--30, 
SLX~1735--269) are clearly seen in the SIGMA image of this region in
1990--1998. Two of them, 1E1740.7--1942 and GRS~1758--258, clearly stand
out by the following properties:

(1) the detection of coherent pulsations or X-ray outbursts from all
other sources provides reliable evidence that the compact object is a
neutron star;

(2) the spectral hardness of all other sources (which is defined as the
ratio of 75--150 to 40--75 keV fluxes) is considerably lower than that for
1E1740.7--2942 and GRS~1758--258 (Goldwurm et al. 1994; Gilfanov et
al. 1995; Churazov et al. 1997).

As was already noted above, these two properties suggest that
1E1740.7--2942 and GRS~1758--258 are black hole candidates. Most of the
binaries that show dynamical evidence for a massive compact object -- a
black hole (apart from a low mass optical companion) are known to be
X-ray transients. Such objects were actually observed by SIGMA in the
Galactic Center region (for example, X-ray Nova Ophiuchi 1993 --
GRS~1716--249; Revnivtsev et al. 1998), with their spectral hardness being
comparable to that of GRS~1758--258 and 1E1740.7--2942. However, it is
clear from the light curves of GRS~1758--258 and 1E1740.7--2942 that
these two sources cannot be classified as classical X-ray transients,
which are characterized by activity periods lasting for several months
against the periods of quiescence (at least tens of years).

Among the other Galactic sources which exhibit fairly hard spectra and
substantial variability on time scales of months and years, we can
mention, for example, Cygnus~X--1, GX~339--4, GRS~1915+105, and GRO~J1655--40\footnote{The last two sources are classified as
transients, although their X-ray activity has been observed for several
years}. Optical observations of GRS~1758--258 rule out the
hypothesis of a very massive companion (Chen et al. 1994; Marti et
al. 1998) similar to the O9.7 optical companion of Cygnus X--1. The
interstellar absorption toward 1E1740.7--2942, $A_{v} > 50$\ (see, e.g., Chen
et al. 1994), is so high that it precludes the possibility
of placing stringent constraints on the optical companion.

Recent observations in $I$\ and $K$\ band (Marti et al. 1998) have revealed
two sources within 1 arcsec of the central radio source which was identified
with GRS~1758--258 (Rodriguez et al. 1992). In the near future, the optical
companion of GRS~1758--258 will be firmly established, which will allow us
to return the problem discussed above and to understand what differs this
source from other X-ray transients.

In their recent papers, Van Paradijs et al. (1996), King et al.
(1997a, 1997b) explain the ``transient'' behavior of most low mass binary
systems with black holes by thermal instability of the accretion disk in
the region of partial hydrogen ionization. This instability was
successfully used to interpret the light curves of cataclysmic variables
(e.g. Meyer and Meyer-Hofmeister 1981). Van Paradijs et al. (1996) point
out that irradiation of the outer parts of the accretion disk by X-ray
emission (Shakura and Sunyaev 1973; Lutyi and Sunyaev 1976) from the
inner zone of main energy release may have a major effect on the
development of this instability for low mass binaries (with a neutron
star or a black hole). Provided that this irradiation is capable of
raising the disk temperature above the hydrogen ionization temperature,
the conditions for the generation of instability vanish and, as a
consequence, the behavior of the accretion rate and the X-ray flux
becomes more regular. In this model, the determining parameter that
separates ``transients'' from ``persistent'' sources is the critical mass
accretion rate at which the outer part of the disk is heated to a
temperature of $\sim$\ 6500K (which corresponds to the hydrogen ionization
temperature). In turn, the radius of the outer disk boundary depends on
the masses of the optical companion and the compact object and on the
orbital period. King et al. (1997b) also note that a distinctive
feature of neutron stars and black holes could be the difference in the
geometry of the emitting region: for black holes, the radiation
originates from the flat inner zone of the accretion disk, while for
neutron stars, a point isotropic source is assumed a the location of the
compact object (as a result, a more appreciable part of the X-ray flux
can be absorbed by the outer parts of the accretion disk around the
neutron star). As a result, a sizable fraction of the low mass binary
systems with neutron stars turn out to be ``persistent'' sources, while
most binary systems with black holes must show up as ``transients''.
Although this model requires further elaboration, it is of interest to
consider the results of its application to GRS~1758--258 and
1E1740.7--2942. The light curves of GRS~1758--258 and 1E1740.7--2942 on
long time scales suggest that the instability under consideration in
these objects is suppressed. This implies that the observed X-ray
luminosity (of the order of several $10^{37}$\ erg/s  at a
distance of 8.5 kpc) heats up the outer  parts of the disk above the hydrogen
ionization temperature. Consequently, (Lutyi \& Sunyaev 1976; Van
Paradijs et al. 1996; King et al. 1997b)

\begin{eqnarray}
T^4_{irr}=\frac{L_X (1-\beta)}{4\pi\sigma R^2} \left( \frac{H}{R}
\right)^{2} \left(\frac{ d \ln H}{d \ln R} - 1 \right) \ga 6500^4
\end{eqnarray}

where $T_{irr}$ is the temperature at the outer disk boundary due to X-ray
irradiation, $L_X$ is the source's X-ray luminosity, $H$ is the disk
half-thickness, $\beta$\ is the disk albedo for X-ray emission, and $R$\  is
the radius of the outer disk boundary. We assume below that $H \propto
R^{9/8}$ (see, e.g., King et al. 1997b), $H/R = 0.2$\ and $\beta = 0.9$ (de
Jong et al. 1996). Consequently,

\begin{eqnarray}
R \la 630 000 L_{X,37}^{1/2} km
\end{eqnarray}

where $L_{X,37}$\ is the luminosity in units of $10^{37}$\ erg/s. According to
Eggleton (1983), the mean Roche-lobe radius (in our case, for the
compact object)

\begin{eqnarray}
{R_{L1}} = a \frac{0.49 q^{2/3}}{0.6 q^{2/3} + ln (1 + q^{1/3})}
\end{eqnarray}

can be determined with an accuracy higher than $\sim$\ 1\% in the interval $0 <
q < \infty$\  (here, $a$\ is the binary separation, $q = M_1/M_2$ is
the mass ratio for the binary system, and $M_1$ and $M_2$ are measured in
solar masses). Using Kepler's third law $P=2\pi \left (
\frac{a^3}{G(M_1+M_2)}\right )^{1/2}$\ and taking $R = 0.7 R_{L1}$
as the radius of the outer disk boundary, we can place the
following constraint on the binary's period:

\begin{eqnarray}
P_h \la 11.8 \frac{(0.6q^{2/3} + ln (1+q^{1/3}))^{3/2}}{q}
\frac{L_{X,37}^{3/4}}{\sqrt{M_1(1+1/q)}}
\end{eqnarray}
where $P_h$\ -- period in hours.

Thus, the conditions for the suppression of instability in a close
binary system with a black hole of mass $M_{1} \ge 3 M_{\sun}$\ and with a
companion 
star of mass $M_{2} \le 4 M_{\sun}$\ (the upper limit for GRS~1758--258;
Chen et al. 
1994) require that the period be shorter than $\sim 7.5$\ h ($L_{X,37}\sim
1$). The assumption of a more massive black hole implies even shorter
periods. Assuming additionally that the optical component is a main sequence
star and that it fills its Roche lobe (assuming that $R_{L2}
\sim \frac{M_{2} R_{\sun}}{M_{\sun}}$), we can 
obtain even more stringent constraints on the period and the mass of the
companion star: $P_h \la 5.8$\ h and $M_2 \la 0.65 M_{\sun}$\ ($M_1 \ga 3.0
M_{\sun}$).

Of course, the above analysis has many significant (and not quite
justified) simplifications and assumptions. Dubus et al. (1998) point
out that due to the substantial optical depth of the disk, the
temperature which is derived from formula (1) reflects only the surface
temperature (as was shown by Lutyi and Sunyaev 1976) and cannot be
directly used to analyze the degree of hydrogen ionization in the bulk
of the disk. Moreover, the increase in the geometrical thickness of the
disk at intermediate distances may result in the self-screening of the outer
regions of the accretion disk. The results of Dubus et al. (1998)
imply appreciably less stringent constraints on the binary's period.
Besides, the additional factor $\frac{H}{R}$, which was used by King (1997a) to
describe the ``disk'' geometry of the emitting region in accreting black
holes, may not be applicable to 1E1740.7--2942 and GRS~1758--258, because
the luminosity is dominated by the hard component that originates in an
optically thin region. The true albedo ($\beta$) may differ
significantly from its assumed empirical value, $\beta=0.9$\, which
is based on observations of low mass X-ray binaries (de Jong et al.
1996). It is, nevertheless, clear that by further analyzing the effect
of irradiation of the outer parts of the accretion disk on the pattern
of accretion, we can place important constraints on the parameters of
binary systems.

Although we assumed above that GRS~1758--258 is not a transient, the source
is undoubtedly variable. The SIGMA observations of GRS~1758--258 in
1990--1998 revealed variations in the hard X-ray flux from the source on time
scales of about one year by more than a factor of 8. During the two
successive series of observations in the Fall of 1991 and in the Spring of
1992, the source flux was below the telescope sensitivity threshold, $\sim
10--15$\ mCrab. Assuming that the source flux was at a comparable level from
September 1991 through April--May 1992, the total time during which the hard
X-ray flux from GRS~1758--258 was low is $\sim$\ 200 days. However, most of
the time -- 14 of the 16 observational series, i.e., $\sim 85\%$\ -- the
source was in a state with a hard X-ray flux above $\sim$\ 30--40 mCrab. A
similar temporal behavior in the hard X-ray range is also observed in the
other two Galactic black hole candidates: 1E1740.7--2942 and Cygnus X--1. As
was already noted above, the luminosity, energy spectra, and the patterns of
aperiodic variability on short time scales for these three sources are also
very similar.

The extended episodes of low hard X-ray luminosity can be attributed
(1) to the transition of the source to a high (soft) spectral state and
(2) to the general decrease in luminosity over the entire X-ray range,
which is caused, for example, by a decrease in the accretion rate. Note
that in the former case, the decrease in the hard X-ray luminosity is
attributable to the redistribution of emitted energy in frequency and
reflects neither the sign nor the amplitude of variations in the
accretion rate. Observations of other sources -- black hole candidates
-- show that the transition to a high spectral state is probably caused
by an increase in the accretion rate by a factor of $\sim 1.5 - 2$.

The RXTE and ASCA observations show (see, e.g., Cui et al. 1997) that for
Cygnus X--1, the extended episodes of low hard X-ray luminosity are
attributable to the transition of the source to a high (soft) spectral
state. For GRS~1758--258, the ROSAT/PSPC flux in March 1992 was half the
measured flux in March 1993 (Grebenev et al. 1996), while the hard X-ray
flux changed by at least a factor of $\approx 5$\ [the 40--150 keV flux did
not exceed 8 mCrab in the Spring of 1992 ($2 \sigma$\ upper limit) and was
$\approx 40$\ mCrab in the Spring of 1993). This may provide evidence that,
as in the case of Cygnus X--1, the extended episode of low hard X-ray
luminosity is attributable at least partly to an appreciable softening of
the source spectrum. At the same time, the decrease in the soft luminosity
does not allow us to assert that the source was in the ``classical'' high
(soft) state in March 1992, similar to that observed in Cygnus X--1.

\section*{Acknowledgments}

This work was supported in part by the INTAS (grant no. 93--3363--ext)
and the Russian Foundation for Basic Research (project no. 96--02--18588).
S.~Kuznetsov was also supported by the ISSEP grants nos. A97--2301 and
A98--1602. We are grateful to S.~Grebenev, A.~King, U.~Kolb, and H.~Ritter
for a discussion and valuable remarks.
\clearpage
\section*{References}

{\small Cordier B., Roques J.P., Churazov E., Gilfanov M., IAUC. 1991, No.
5377.}

{\small Chen W., Gehrels N., Leventhal M., ApJ, 1994, 426, 586.}

{\small Churazov E., Gilfanov M., Sunyaev R. et al., Adv. Space Res., 1997,
19, 61.} 

{\small Cui W., Zhang S.N., Focke W., Swank J.H., ApJ, 1997, 484, 383.}

{\small de Jong J.A., Van Paradijs J., Augusteijn T., A\&A, 1996, 314, 484.}

{\small Dubus G., Lasota J.P., Hameury J.M., Charle P., MNRAS, 1998, (in
press).}

{\small Eggleton P., ApJ, 1983, 268, 368.}

{\small Gilfanov M., Churazov E., Sunyaev R., Khavenson N.,Novikov B.,
Dyachkov A., Kremnev R., Sukhanov K., Bouchet L., Mandrou P.,Roques
J.P., Vedrenne G., Cordier B., Goldwurm A., Laurent P., Paul
J., ApJ, 1993, 418, 844.}

{\small Gilfanov M., Churazov E., Sunyaev R., Vikhlinin A.,Finoguenov A.,
Sitdikov A., Dyachkov A., Khavensov N., Laurent P., Ballet J.,Claret
A.,Goldwurm A., Roques J.P., Mandrou P., Niel M., Vedrenne G.,
Lives of the the Neutron Stars. NATO ASI Ser. (ed. Alpar
M.,Kizilo\v{g}lu \"U., Van Paradijs J.), Dordrecht: Kluwer Acad. Publ.,
1995, 450, 712.}

{\small Grebenev A., Pavlinsky M., Sunyaev R.,
The Transparent Universe, 2nd Integral Workshop (ed. Winkler C.,
Courvoisier T., Durouchoux P., Kaldeich-Sch\"urmann B.) Noordwijk: ESA Publ.
Div. ESTEC, 1997, 382, 183.}

{\small Goldwurm A., Cordier B., Paul J., Ballet J., Bouchet L., Roques
J.P., Vedrenne G., Mandrou P., Sunyaev R., Churazov E., Gilfanov M.,
Finoguenov A., Vikhlinin A., Dyachkov A., Khavenson N., Kovtunenko V.,
Nature, 1994, 371, 589.}

{\small Kelloggg E., Baldwin J., Koch D., ApJ, 1975, 199, 299.}

{\small King A., Frank J., Kolb U., Ritter H., ApJ, 1997a, 484, 844.}

{\small King A., Kolb U., Szuszkiewicz E., ApJ, 1997b, 488, 89.}

{\small Kuznetsov S., Gilfanov M., Churazov E., Sunyaev R.,Korel I.,
Khavenson N., Dyachkov A., Chulkov I., Ballet J., Laurent P., Vargas M.,
Goldwurm A., Roques J.P., Jourdain E., Bouchet L., Borrel V.,
MNRAS, 1997, 292, 651.}

{\small Lutyj V., Sunyaev R., Astronomicheskij zhurnal, 1976, 53, 511.}

{\small Marti J., Mereghetti S., Chaty S., Mirabel I., Goldoni P.,
Rodriguez L., A\&A, 1998, 338, L95.}

{\small Mereghetti S., Caraveo P., Bignami G.F., Belloni T., A\&A, 1992,
259, 205.}

{\small Mereghetti S., Belloni T., Goldwurm A., ApJ, 1994, 433, L21.}

{\small Meyer F., Meyer-Hofmeister E., A\&A, 1981, 104, 10.}

{\small Mirabel F., Rodriguez L.F., Cordier B., Paul J., Lebrun F.,
Nature, 1992, 358, 215.}

{\small Mirabel F., Rodriguez L.F., Cordier B., Paul J., Lebrun F., A\&ASS,
1993, 97, 193.}

{\small Miyamoto S., Kitamoto, S., Mitsuda K., Dotani, T., ApJ, 1992,
391, L21.}

{\small Paul J., Ballet J., Cantin M., Cordier B., Goldwurm A.,Lambert A.,
Mandrou P., Chabaud J.P., Ehanno M., Lande J., Adv. Space Res., 1991, 11, 289.}

{\small Revnivtsev M., Gilfanov M., Churazov E., Sunyaev R.,Borozdin K.,
Aleksandrovich N., Khavenson N., Chulkov I., Goldwurm A., Ballet J.,
Denis M., Laurent P., Roques J.P., Borrel V., Bouchet L., Jourdain E.,
A\&A, 1998, 331, 557.}

{\small Rodriguez L.F., Mirabel I.F., Marti J., ApJ, 1992, 401, L15.}

{\small Shakura N.I., Sunyaev R.A, A\&A, 1973, 24, 337.}

{\small Shapiro S., Lightman A., Eardley D., ApJ, 1976, 204, 187.}

{\small Smith D.M., Heindl W.A., Swank J., Leventhal M., MirabelI.,
Rodriguez L., ApJ, 1997, 489, L51.}

{\small Sunyaev R.A., Titarchuk L.G., A\&A, 1980, 86, 121.}

{\small Sunyaev R.A., Gilfanov M.R., Churazov E.M., Pavlinsky M.N.,
Babalyan G., Dekhanov I., Kuznetsov A., Grebenev S., Yunin S., Yamburenko N.,
Cordier B., Lebrun F., Laurent P., Ballet J., Mandrou P., Roques J.-P.,
Vedrenne G., Boucher L., Astron. Lett., 1991, 17, 116.}

{\small Van Paradijs J., ApJ,1996, 464, 139.}


\begin{deluxetable}{lccc}
\tablenum{1. Power-law best-fit parameters for GRS~1758--258 spectra in the
40--300 keV energy band}
\small
\tablecolumns{4}
\tablehead{
\colhead{Date of observations}&
\colhead{Photon index ($\alpha$)}&
\colhead{Luminosity,}&
\colhead{$\chi^{2}$(d.o.f.)}\nl
\colhead{~}&
\colhead{~}&
\colhead{(40--200 keV)\tablenotemark{a}}&
\colhead{~}
}
\startdata
\cutinhead{\it 1990}
Spring 	& $1.91\pm 0.17$ & 11.2   & 29.8(44)\nl
Fall 	& $2.18\pm 0.09$ & 11.6   & 44.0(44)\nl
Mean & $2.13\pm 0.08$ & 11.5   & 37.9(44)\nl
\cutinhead{\it 1991}
Spring 	& $2.17\pm 0.16$ & 8.6    & 50.1(44)\nl
\cutinhead{\it 1992}
Fall 	& $2.07\pm 0.44$ & 4.8    & 49.9(44)\nl
\cutinhead{\it 1993}
Spring 	& $1.82\pm 0.22$ & 5.9    & 58.7(44)\nl
Fall 	& $1.94\pm 0.18$ & 9.7    & 45.3(44)\nl
Mean & $1.88\pm 0.14$ & 7.4    & 37.5(44)\nl
\cutinhead{\it 1994}
Spring 	& $2.21\pm 0.15$ & 10.6   & 48.9(44)\nl
Fall 	& $2.13\pm 0.29$ & 5.5    & 45.0(44)\nl
Mean & $2.19\pm 0.14$ & 8.6    & 51.3(44)\nl
\cutinhead{\it 1995}
Fall 	& $2.25\pm 0.45$ & 7.8    & 43.7(44)\nl
\cutinhead{\it 1996}
Spring 	& $2.14\pm 0.49$ & 6.7    & 55.0(44)\nl
\enddata
\tablenotetext{a}{ $\times 10^{36}$\ Erg s$^{-1}$.}
\end{deluxetable}


\clearpage

\begin{deluxetable}{lccc}
\tablenum{2. Optically-thin thermal bremsstrahlung model best-fit parameters
for GRS~1758--258 spectra in the 40--300 keV energy band (Kellog et al. 1975)}
\small
\tablecolumns{4}
\tablehead{
\colhead{Date of observations}&
\colhead{Temperature ($kT$),}&
\colhead{Luminosity,}&
\colhead{$\chi^{2}$(d.o.f.)}\nl
\colhead{~}&
\colhead{keV}&
\colhead{(40--200 keV)\tablenotemark{a}}&
\colhead{~}
}
\startdata
\cutinhead{\it 1990}
Spring   & $177^{+76}_{-46}$ & 11.3   & 28.4(44)\nl
Fall 	& $111^{+15}_{-12}$ & 11.6   & 40.5(44)\nl
Mean & $121^{+16}_{-13}$ & 11.6   & 32.8(44)\nl
\cutinhead{\it 1991}
Spring 	& $110^{+30}_{-22}$ & 8.6    & 46.6(44)\nl
\cutinhead{\it 1992}         
Fall 	& $131^{+161}_{-50}$ & 4.9    & 48.7(44)\nl
\cutinhead{\it 1993}
Spring 	& $234^{+186}_{-82}$ & 6.0    & 59.1(44)\nl
Fall 	& $182^{+93}_{-52}$ & 9.8    & 47.9(44)\nl
Mean & $206^{+82}_{-50}$ & 7.5    & 39.9(44)\nl
\cutinhead{\it 1994}
Spring 	& $111^{+28}_{-22}$ & 10.6   & 49.2(44)\nl
Fall 	& $118^{+64}_{-38}$ & 5.6    & 42.3(44)\nl
Mean & $113^{+26}_{-19}$ & 8.6    & 47.4(44)\nl
\cutinhead{\it 1995}
Fall 	& $108^{+111}_{-44}$ & 7.7    & 43.9(44)\nl
\cutinhead{\it 1996}
Spring 	& $130^{+180}_{-59}$ & 6.8    & 54.6(44)\nl
\enddata
\tablenotetext{a}{ $\times 10^{36}$\ Erg s$^{-1}$.}
\end{deluxetable}


\clearpage

\begin{deluxetable}{lcccc}
\tablenum{3. Comptonized disc model best-fit parameters for the spectra of
GRS~1758--258 in the 40--300 keV energy band (Sunyaev, Titarchuk 1980)}
\small
\tablecolumns{5}
\tablehead{
\colhead{Date of observations}&
\colhead{Temperature ($kT$),}&
\colhead{Optical} &
\colhead{Luminosity,}&
\colhead{$\chi^{2}$(d.o.f.)}\nl
\colhead{~}&
\colhead{keV}&
\colhead{depth ($\tau$)}&
\colhead{(40--200 keV)\tablenotemark{a}}&
\colhead{~}
}
\startdata
\cutinhead{\it 1990}
Spring 	&$43^{+45}_{-13}$ &$1.3^{+1.7}_{-0.8}$ & 11.4   & 28.4(43)\nl
Fall 	&$45^{+20}_{-9}$ &$0.9^{+0.3}_{-0.4}$ & 11.7   & 40.4(43)\nl
Mean &$45^{+15}_{-10}$ &1.0$^{+0.3}_{-0.3}$ & 11.6   & 32.9(43)\nl
\cutinhead{\it 1991}
Spring 	&$31^{+11}_{-9}$ &$1.6^{+2.2}_{-0.6}$ & 8.4    & 45.7(43)\nl
\cutinhead{\it 1992}
Fall 	&$34^{+14}_{-34}$ &$1.9^{+\infty}_{-0.9}$ & 5.5    & 47.2(43)\nl
\cutinhead{\it 1993}
Spring 	&$85^{+\infty}_{-37}$ &$0.6^{+0.7}_{-0.6}$ & 6.1    & 58.6(43)\nl
Fall 	&$127^{+\infty}_{-50}$ &$0.3^{+0.3}_{-0.3}$ & 9.8    & 45.6(43)\nl
Mean &$146^{+\infty}_{-73}$ &$0.3^{+0.4}_{-0.3}$ & 7.4    & 37.6(43)\nl
\cutinhead{\it 1994}
Spring 	&$68^{+\infty}_{-38}$ &$0.5^{+0.6}_{-0.5}$ & 10.7   & 48.6(43)\nl
Fall 	&$25^{+12}_{-25}$ &$4.6^{+\infty}_{-2.7}$ & 5.6    & 39.0(43)\nl
Mean	&$37^{+13}_{-7}$ &$1.2^{+0.6}_{-0.5}$ & 8.6    & 46.7(43)\nl
\cutinhead{\it 1995}
Fall 	&$82^{+\infty}_{-58}$ &$0.4^{+2.2}_{-0.4}$ & 7.8    & 43.7(43)\nl
\cutinhead{\it 1996}
Spring 	&$39^{+\infty}_{-16}$ &$1.3^{+5.3}_{-1.3}$ & 7.7    & 54.4(43)\nl
\enddata
\tablenotetext{a}{ $\times 10^{36}$\ Erg s$^{-1}$.}
\end{deluxetable}


\clearpage

\begin{deluxetable}{lcc}
\tablenum{4. Best-fit  parameters for the spectra of GRS~1758--258 and
1E1740.7--2942 in the 40--300 keV energy band averaged over all GRANAT/SIGMA
observations in 1990-1997}
\small
\tablecolumns{3}
\tablehead{
\colhead{Parameter}&
\colhead{GRS~1758--258}&
\colhead{1E1740.7--2942}\nl
}
\startdata
\cutinhead{\it Power-law}
Slope ($\alpha$) & $2.06\pm 0.06$ & $2.03\pm 0.03$ \nl
Luminosity (40--200 keV)\tablenotemark{a} & $8.0\pm 0.2$ & $10.1\pm 0.2$\nl
$\chi^{2}$(d.o.f.) & 59.9(44) & 70.1(44) \nl
\cutinhead{\it Power-law with high energy exponential cutoff}
Slope ($\alpha$) & $1.0\pm 0.3$ & $1.2\pm 0.2$ \nl
cutoff energy (keV) & $89^{+40}_{-20}$ & $110^{+26}_{-23}$ \nl
Luminosity (40--200 keV)\tablenotemark{a} & $8.0\pm 0.2$ & $10.2\pm 0.2$ \nl
$\chi^{2}$(d.o.f.) & 45.1(43) & 40.1(43)\nl
\cutinhead{\it Optically-thin thermal bremsstrahlung\tablenotemark{b}}
Temperature, keV & $135\pm 12$ & $142\pm 8$ \nl
Luminosity (40--200 keV)\tablenotemark{a} & $8.0\pm 0.2$ & $10.2\pm 0.2$ \nl
$\chi^{2}$(d.o.f.) & 46.0(44) & 40.9(44) \nl
\cutinhead{\it Comptonized disc model\tablenotemark{c}}
Temperature, keV & $41^{+7}_{-5}$ & $44^{+5}_{-4}$ \nl
Optical depth ($\tau$) & $1.2\pm 0.2$ & $1.1\pm 0.1$ \nl
Luminosity (40--200 keV)\tablenotemark{a} & $8.0\pm 0.2$ & $10.2\pm 0.2$ \nl
$\chi^{2}$(d.o.f.) & 44.9(43) & 43.2(43) \nl
\enddata
\tablenotetext{a}{ $\times 10^{36}$\ Erg s$^{-1}$.}
\tablenotetext{b}{ Kellogg et al. (1975)}
\tablenotetext{c}{ Sunyaev, Titarchuk (1980)}

\end{deluxetable}

\clearpage

\begin{figure}
\figurenum{1}
\plotone{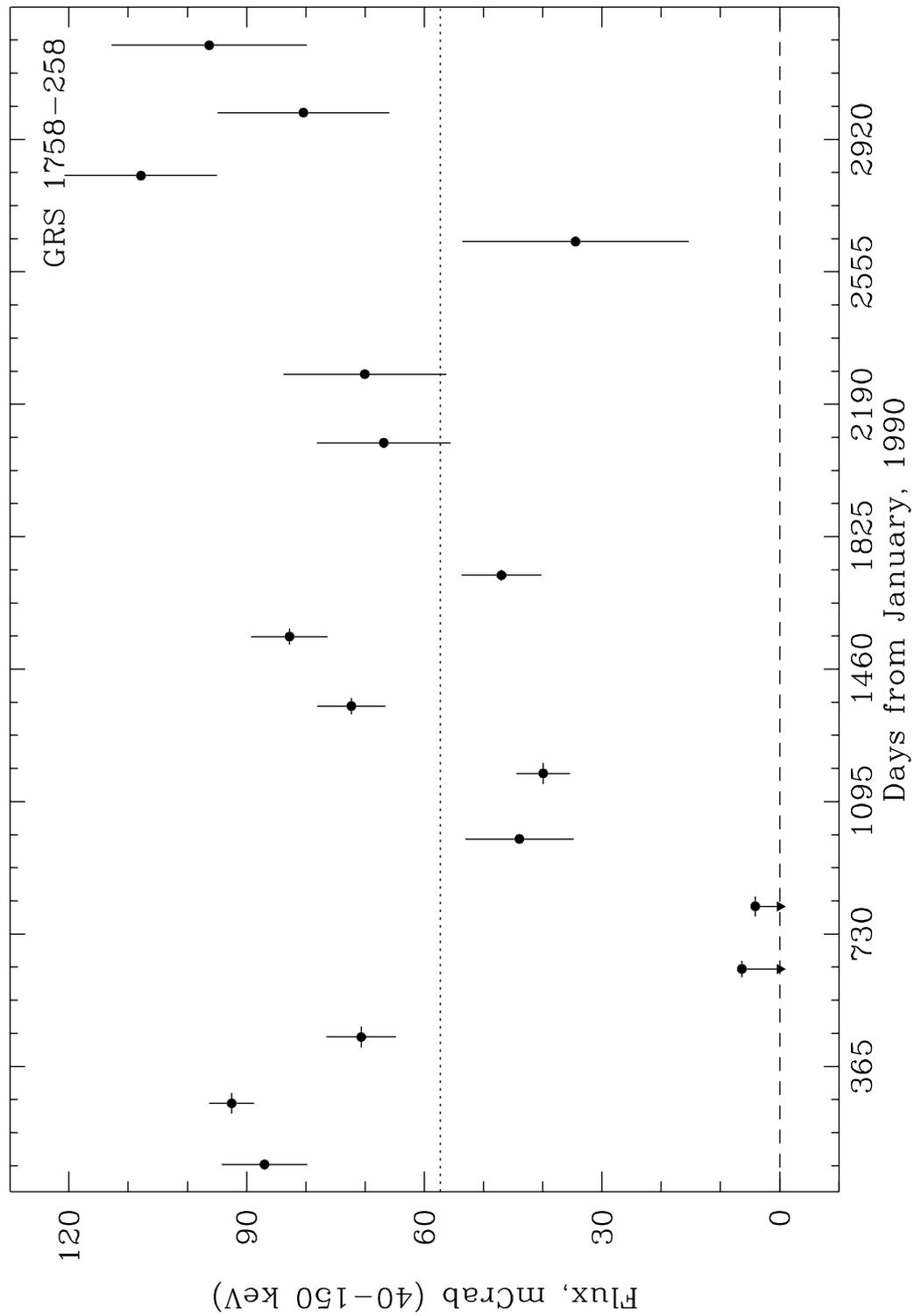}
\epsscale{1.0}
\caption{Light curve of GRS~1758--258 in the band 40--150 keV obtained
during the GRANAT/SIGMA observations in 1990-1998. The flux averaged
over observational sets is indicated by the filled circles. The mean
source intensity is indicated by the dotted line. January 1, 1990,
corresponds to MJD 47892.}

\end{figure}

\clearpage

\begin{figure}
\figurenum{2}
\plotone{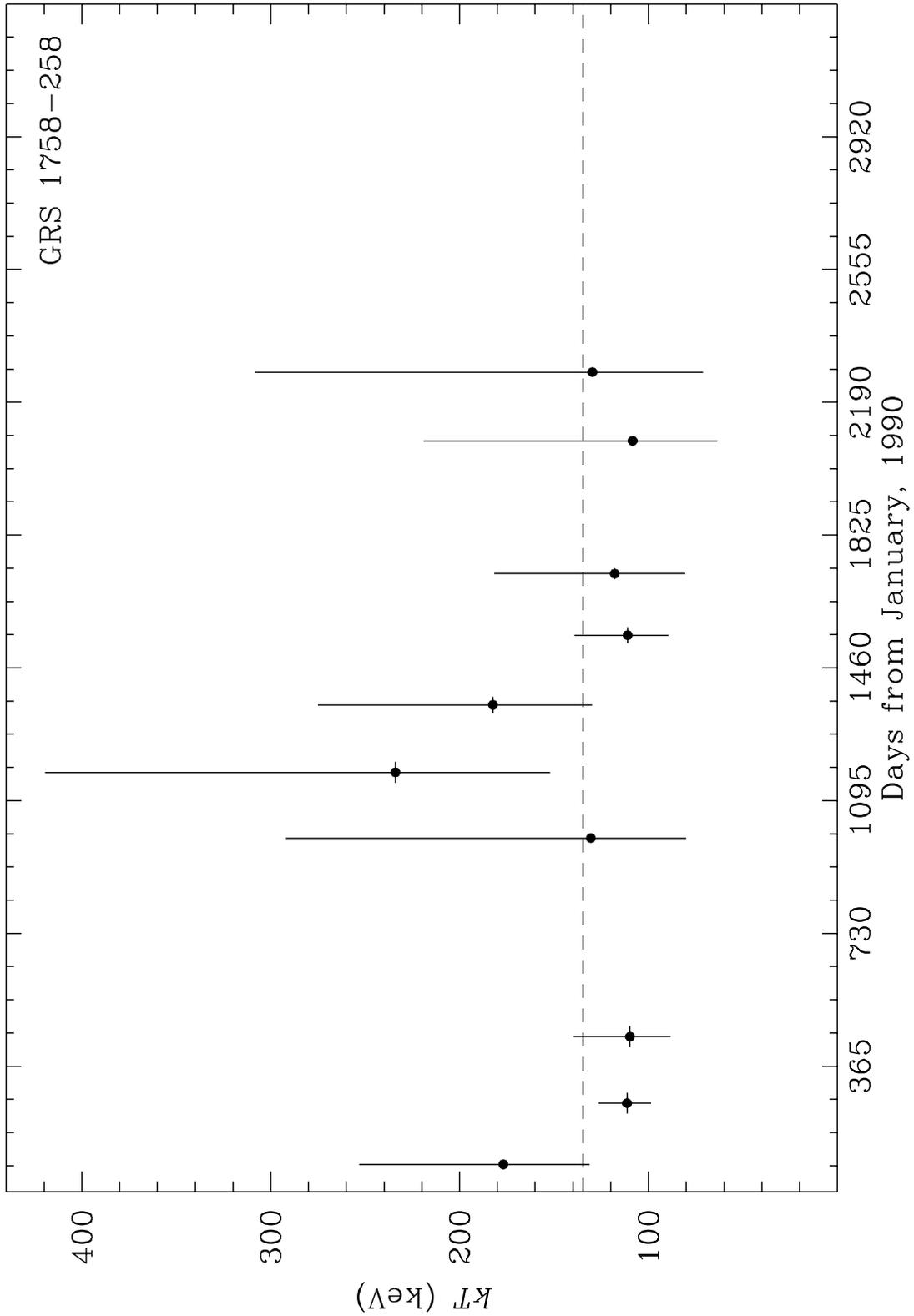}
\epsscale{1.0}
\caption{Time evolution of the spectral hardness of GRS~1758--258 in the
band 40-300 keV. The hardness is characterized by the best-fit
bremsstrahlung temperature (Kellogg et al. 1975); see Table 2 for
details. The data were averaged over observational sets. January 1, 1990
corresponds to MJD 47892. The best-fit bremsstrahlung temperature of the
source spectrum averaged over all SIGMA observations (1990-1997) is
indicated by the dashed line.}

\end{figure}

\clearpage

\begin{figure}
\figurenum{3}
\plotone{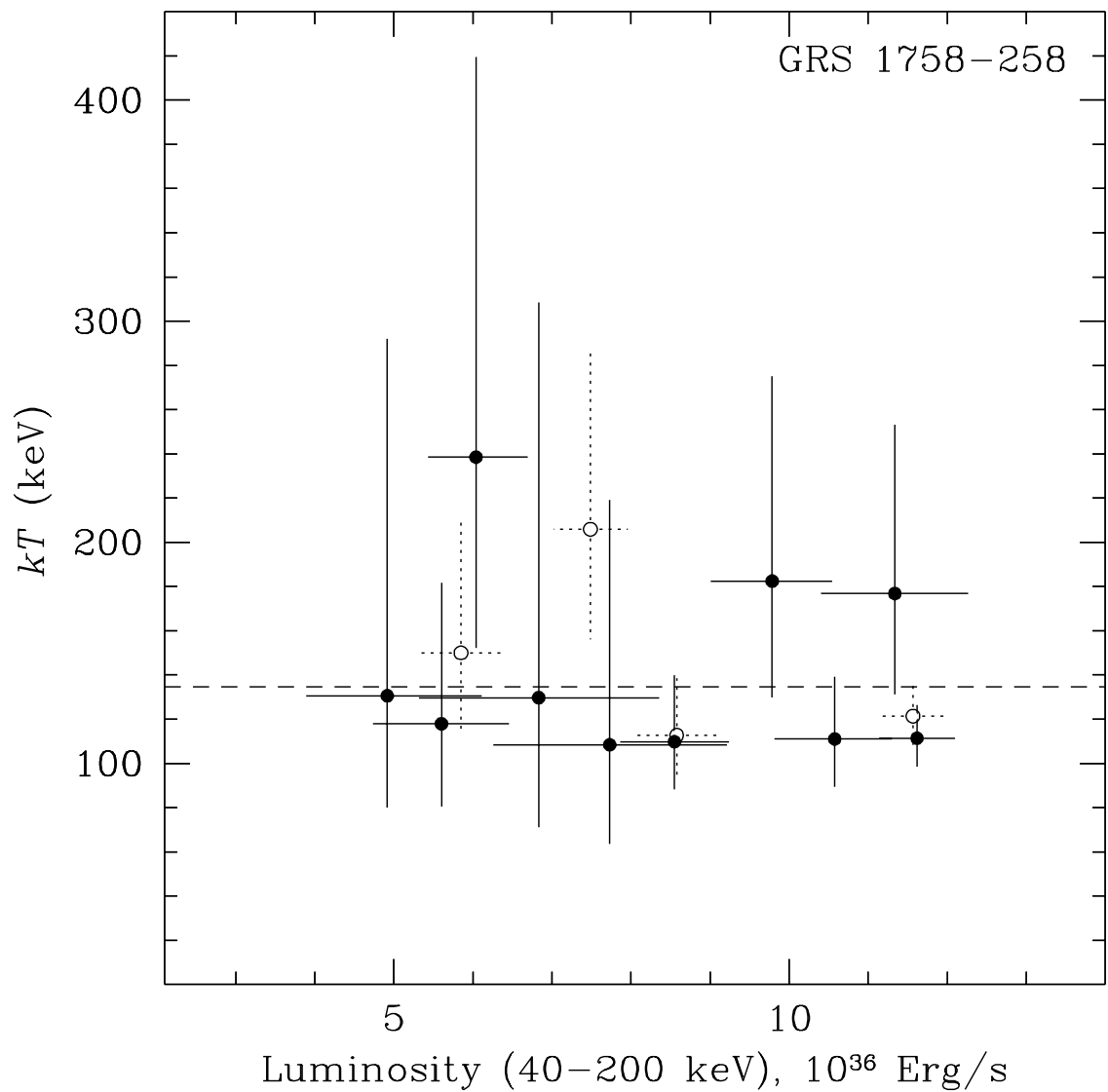}
\epsscale{1.0}
\caption{Same as Fig. 2 for the spectra averaged over observational sets
(filled circles) and over years (open circles).}

\end{figure}

\clearpage

\begin{figure}
\figurenum{4}
\plotone{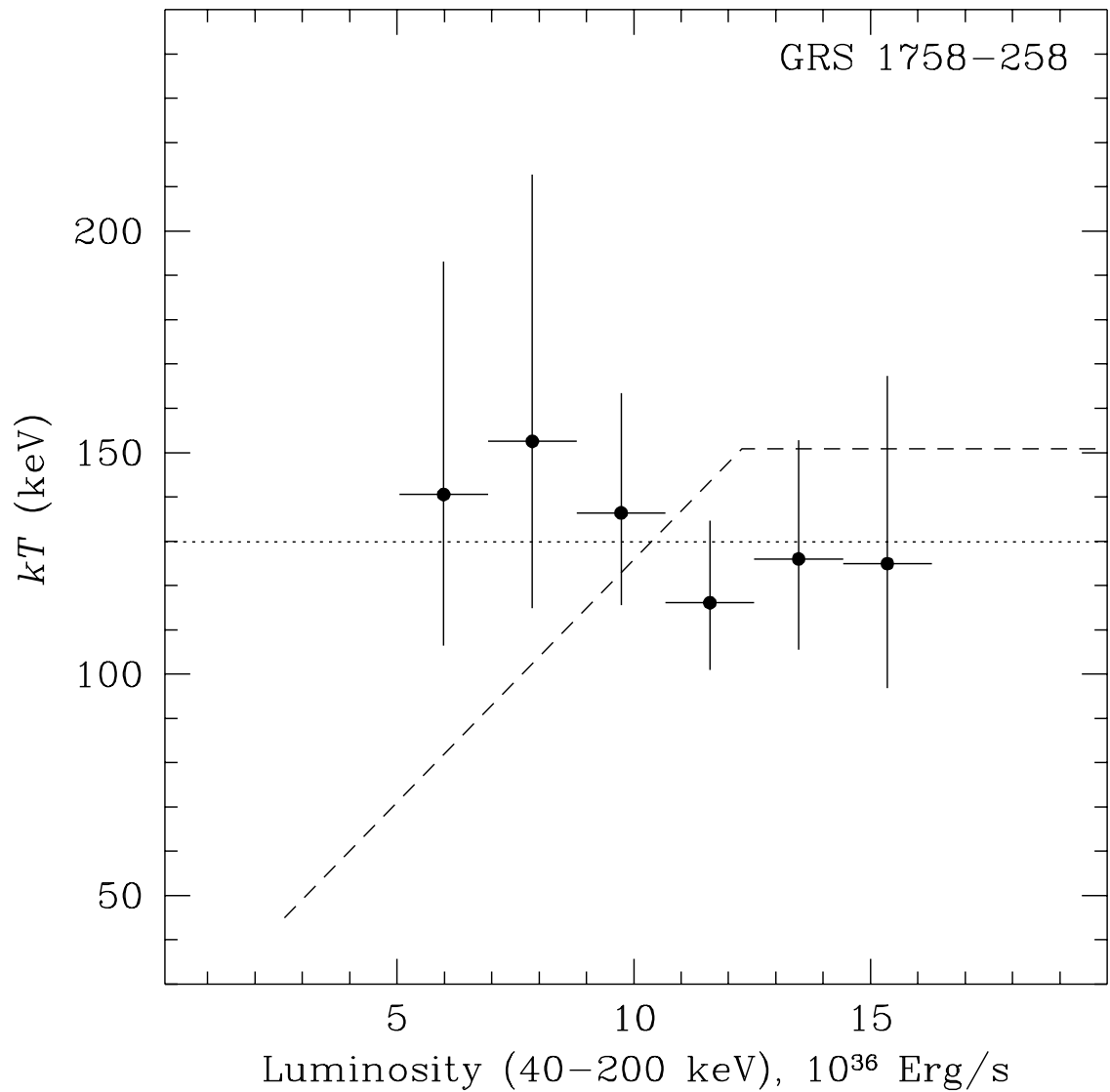}
\epsscale{1.0}
\caption{Spectral hardness of the source (see the caption to Fig. 2)
versus X-ray luminosity. The hardness averaged over all points is
indicated by the dotted line. The dependence for Cyg X-1 and
1E1740.7--2942 (Kuznetsov et al. 1997) is indicated by the dashed line.}
\end{figure}
\clearpage

\begin{figure}
\figurenum{5}
\plotone{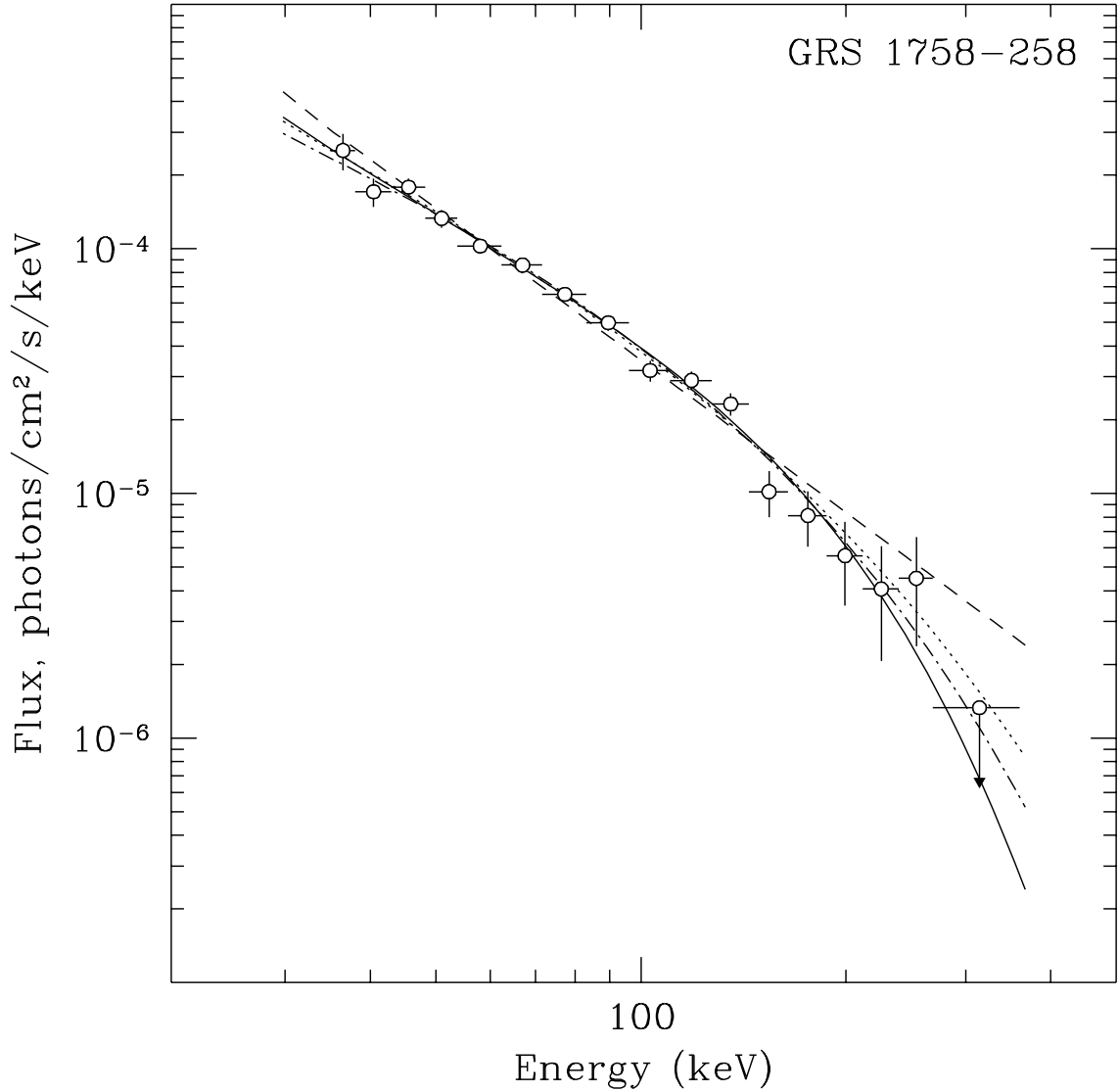}
\epsscale{1.0}
\caption{Hard X-ray spectrum of GRS~1758--258 averaged over all SIGMA
observations in 1990-1997. The best-fit model spectra are indicated by
the lines (see Table 4 for details): the solid line for a comptonized
disk, the dotted line for optically thin thermal bremsstrahlung, the
dashed line for a power law, and the dash-dotted line for a power law
with an exponential cutoff at high energies.}

\end{figure}

\end{document}